\documentclass[aps,pra,twocolumn,groupedaddress,amsmath,amssymb,superscriptaddress]{revtex4-1}
\usepackage{graphicx}  
\usepackage{dcolumn}   
\usepackage{bm}        
\usepackage{verbatim}   
\usepackage{upgreek}   
\usepackage{hyperref}
\usepackage{braket}
\usepackage{color}

\begin{document}

\title{Rydberg Electrometry for Optical Lattice Clocks}

\begin{abstract}
Electrometry is performed using Rydberg states to evaluate the quadratic Stark shift of the $5s^2$~$^1$S$_0-5s5p$~$^3$P$_0$ clock transition in strontium. By measuring the Stark shift of the highly excited $5s75d$~$^1$D$_2$ state using electromagnetically induced transparency, we characterize the electric field with sufficient precision to provide tight constraints on the systematic shift to the clock transition. Using the theoretically derived, and experimentally verified, polarizability for this Rydberg state we can measure the residual field with an uncertainty well below 1 $\textrm{V} \textrm{m}^{-1}$. This resolution allows us to constrain the fractional frequency uncertainty of the quadratic Stark shift of the clock transition to $2\times10^{-20}$. 
\end{abstract}

\pacs{xxx}

\author{W.~Bowden}
\email[]{william.bowden@npl.co.uk}
\affiliation{National Physical Laboratory, Teddington, Middlesex, UK }
\affiliation{Department of Physics, University of Oxford, Oxford, Oxfordshire, UK }

\author{R.~Hobson}
\affiliation{National Physical Laboratory, Teddington, Middlesex, UK }

\author{P.~Huillery}
\affiliation{Joint Quantum Centre Durham-Newcastle, Department of Physics, Durham University, Durham UK }
%

\author{P.~Gill}
\affiliation{National Physical Laboratory, Teddington, Middlesex, UK }

\author{M.~P.~A~Jones}
\affiliation{Joint Quantum Centre Durham-Newcastle, Department of Physics, Durham University, Durham UK }

\author{I.~R.~Hill}
\affiliation{National Physical Laboratory, Teddington, Middlesex, UK }

\maketitle 
\section{Introduction}

In the past decade, optical lattice clocks \cite{Ludlow2015,Katori2011} have made dramatic progress in accuracy and stability, surpassing their microwave counterparts to have the lowest fractional uncertainty of any frequency standard to date. Ensuring the continuation of this progress demands that the environmental perturbations affecting their accuracy are characterized to increasingly precise levels. Motivated by this challenge, we report on a new method using highly excited Rydberg states to provide an \textit{in situ} measurement of the DC electric field.

Uncharacterized electric fields can severely impact the accuracy of an atomic clock. For the $5s^2$~$^1$S$_0-5s5p$~$^3$P$_0$ clock transition in strontium, an electric field of 570 V/m yields a DC Stark shift of 1 Hz \cite{Middelmann2012}, or $2\times10^{-15}$ in fractional units, some three orders of magnitude above the lowest estimated total inaccuracy of a strontium optical lattice clock \cite{Nicholson2015}. Where dielectric surfaces are close to the atoms, shifts as large as 1$\times 10^{-13}$ have been observed \cite{Lodewyck2012}. While steps can be taken to reduce the residual electric field seen by the reference atoms, such as Faraday shielding \cite{Beloy2014}, or UV discharge of dielectric surfaces \cite{Pollack2010}, a characterization of the remaining field is necessary. This is typically done by direct spectroscopy of the clock transition with an externally applied electric field. With no residual electric field present, the quadratic nature of the perturbation implies the resulting induced frequency shift should be unchanged if the polarity of the applied field is reversed but the magnitude is left unchanged \cite{Matveev2011}. However, this method relies on the ability to apply sufficiently large and stable electric fields at the position of the atoms to induce a shift large enough to be quickly resolved during operation of the clock. For metallic vacuum chambers with minimal dielectric openings and no internal electrodes, producing such a shift is problematic. Furthermore, the applied field can charge dielectric materials such as the vacuum viewports \cite{Abel2011}, resulting in a time dependence of the effective applied field.

We circumvent these challenges by performing \emph{in-situ} electrometry using electromagnetically induced transparency (EIT) spectroscopy \cite{Fleischhauer2005} to measure the quadratic Stark shift of the Sr $5s75d$~$^1D_2$~$m_J = 0,\pm1,\pm2$ Rydberg states. Rydberg states of alkaline earth atoms are of growing interest for applications in quantum information \cite{Daley2008} and many body physics  \cite{Mukherjee2011}, motivating their study by several groups \cite{Millen2010,DeSalvo2016}. The low-frequency polarizability scales with principal quantum number $n$ as $n^7$, making Rydberg states well suited for AC \cite{Sedlacek2012,Holloway2014,Fan2015} and DC \cite{Osterwalder1999, Thiele2015, Doughty1984} electrometry, with EIT spectroscopy being a particularly convenient measurement technique \cite{Abel2011, Mohapatra2007, Mohapatra2008,Tauschinsky2010}. The polarizability of our chosen Rydberg state is eight orders of magnitude larger than that of the clock transition, which reduces the required spectroscopic resolution from sub Hz, as needed when using to clock transition, to MHz when using Rydberg states to achieve the desired level of inaccuracy. It has also been proposed to use Rydberg states to measure ambient black-body radiation \cite{Ovsiannikov2011} which is responsible for the leading systematic uncertainty in many current Sr lattice clocks \cite{LeTargat2013, Falke2014, Poli2014, Takamoto2005}. 

Using this spectroscopic method, we reduce the fractional uncertainty of the DC Stark shift of the clock transition to $2\times10^{-20}$. Furthermore the formation of Rydberg states in a system designed for the operation as an atomic clock opens the possibility to investigate proposals to use long range Rydberg interactions to generate squeezed states which exhibit reduced quantum projection noise \cite{Gil2014}.

\section{Theory: Single electron model and Stark maps }
\label{theory}
Alongside their large polarizability, another key advantage of Rydberg states for precision electrometry is that their Stark map - the variation of energy levels with the applied electric field - may be calculated to a very high degree of accuracy \cite{Zimmerman1979}. Even in divalent atoms such as strontium, where inter-electronic Coulomb interactions lead to perturbations of the Rydberg states \cite{Gallagher1994}, it can be shown that accurate wavefunctions \cite{Vaillant2012,Vaillant2014,Ye2012}, and Stark maps \cite{Millen2011,Lochead2013,Hiller2014} can be obtained without recourse to the complex atomic structure calculations required for the clock states \cite{Safronova2013}.

\begin{figure}[h]
    \centering
    \includegraphics[width=0.45\textwidth]{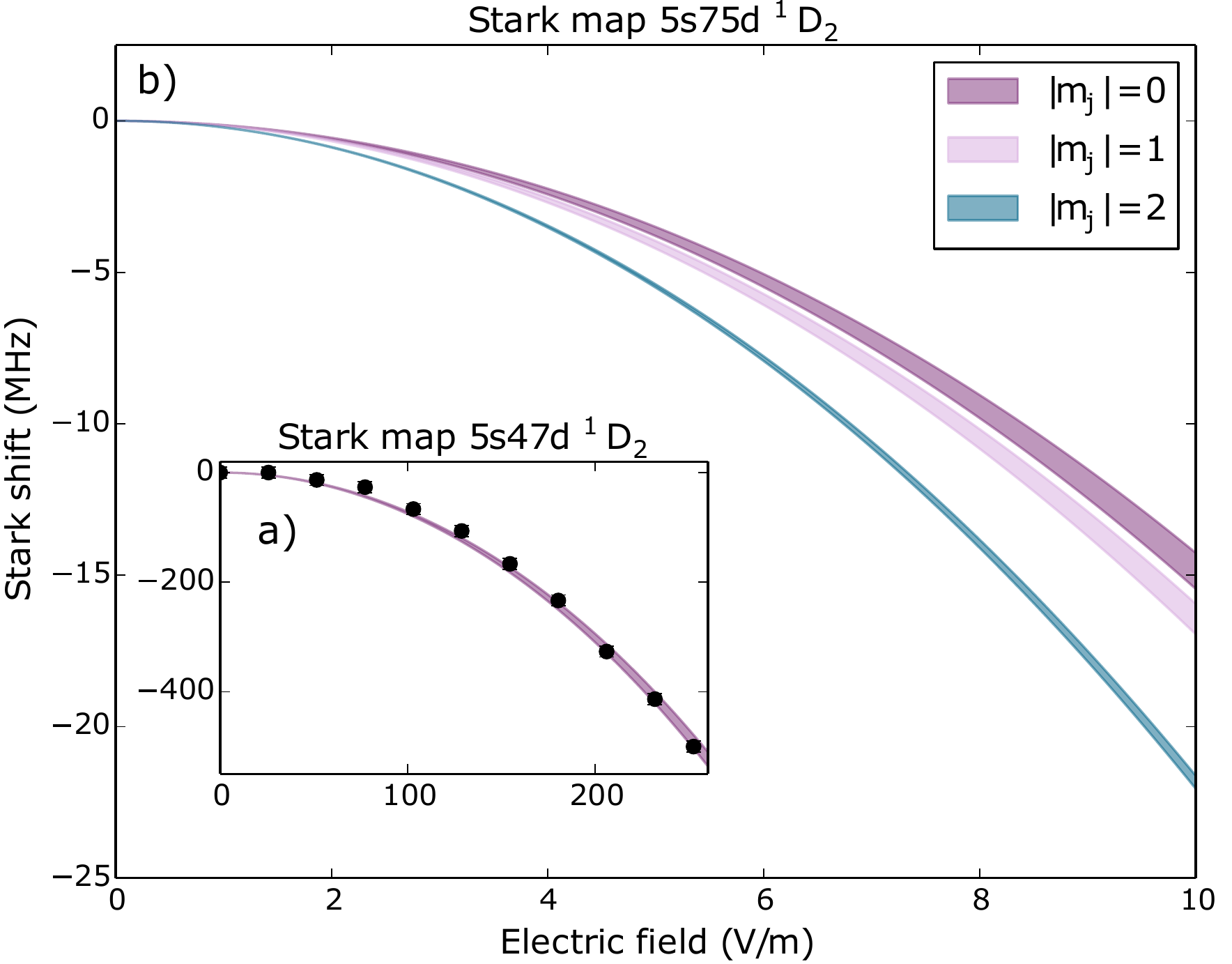}
    \caption{\label{fig:fig1} (a) Predicted Stark shift of the  $5s47d$~$^1$D$_2$ $m_J$ = 0 state compared to experimental data (black dots) without any adjustable parameters. (b) Predicted Stark shift of the three $\vert m_J \vert$ components of the $5s75d$~$^1$D$_2$ state. The shaded area represents the uncertainty arising from the experimental determination of the zero-field energy (see text).}
\end{figure}
This simplification occurs because for Rydberg states, the effect of interelectronic interactions occurs primarily through the existence of spatially compact doubly-excited perturber states that overlap in energy with the Rydberg manifold. However since the static polarizability is dominated by the long-range character of the wavefunction \cite{Vaillant2015}, these states do not significantly alter the Stark maps. In previous work we have shown that an effective one-electron treatment that neglects inter-electronic effects gives Stark maps that are agreement with measurements for high-lying strontium Rydberg states \cite{Millen2011}.
	
	 Here we develop this approach to calculate Stark maps with the well-characterized uncertainty necessary to constrain the instability due to the electric field. The method is based on analytic expressions for the wavefunctions and dipole matrix elements generated using the Coulomb approximation \cite{Zimmerman1979}. The wavefunction is parametrized by a quantum defect, which is obtained by fitting to the experimentally measured zero-field energies.
	To obtain the Stark map, Rydberg states within a range of $[n-3,n+5]$ from the target states and with $l \in [0,15]$ are included in the Stark Hamiltonian, which is then diagonalised numerically for each value of the field. At low electric field, the Stark shift of non-degenerate states is approximately quadratic, and a fit of the form $\Delta_\mathrm{E} = \frac{1}{2}\alpha_0 E^2$ yields the static polarizability $\alpha_0$. 
	
 In Fig.\ref{fig:fig1}(a) we show an example Stark map compared to experimental measurements taken in an atomic beam apparatus with a well-defined electrode geometry \cite{Hanley2017}. The data and the model are in quantitative agreement without any adjustable parameters. The predicted Stark map for the higher-lying state used to constrain the field in the lattice clock is shown in Fig.\ref{fig:fig1}(b). In both cases, the shaded band indicates the theoretical uncertainty in the Stark map. By far the dominant contribution to this uncertainty is the experimental uncertainty in the zero-field energies used to calculate the wavefunctions. In strontium, the current state-of-the-art absolute frequency measurements of Rydberg energy levels is $\pm30$ MHz for $S$ and $D$ states \cite{Beigang1982}, with measurements on the other series having much greater errors \cite{Rubbmark1978}. The zero-field energies for each series and the corresponding errors are obtained by fitting these experimental data with the Rydberg-Ritz formula \cite{Vaillant2012}. The shaded region corresponds to the extremal cases where the 1-sigma errors on each series are combined to give the extremal overall polarizabilities. Using ultracold atoms and frequency comb technology, it was recently shown that absolute Rydberg spectroscopy with 10 kHz uncertainty is possible \cite{Kliese2016}, opening the way to significant improvement in the Stark map uncertainty.

\section{ Experimental Approach }

We follow the standard approach for producing cold strontium samples \cite{katori1999, Loftus2004, Sorrentino2006}. We operate a blue magneto-optical trap (MOT) on the 461 nm transition for 650 ms followed by a broadband and single-frequency red MOT for 100 ms and 150 ms, respectively,  which results in a sample of approximately $10^5$ $^{88}$Sr atoms at a temperature of around 1 $\upmu$K. Details of our apparatus can be found in \cite{Hill2014} and \cite{Hill2016}. Before implementing the EIT probe pulse, the cloud of atoms is released from the red MOT for 5 ms, giving time for the magnetic field to settle to the desired bias value, and for the atoms to expand ballistically to a lower density which was observed to improve the signal to noise level. 

\begin{figure}[h]
    \centering
    \includegraphics[width=0.47\textwidth]{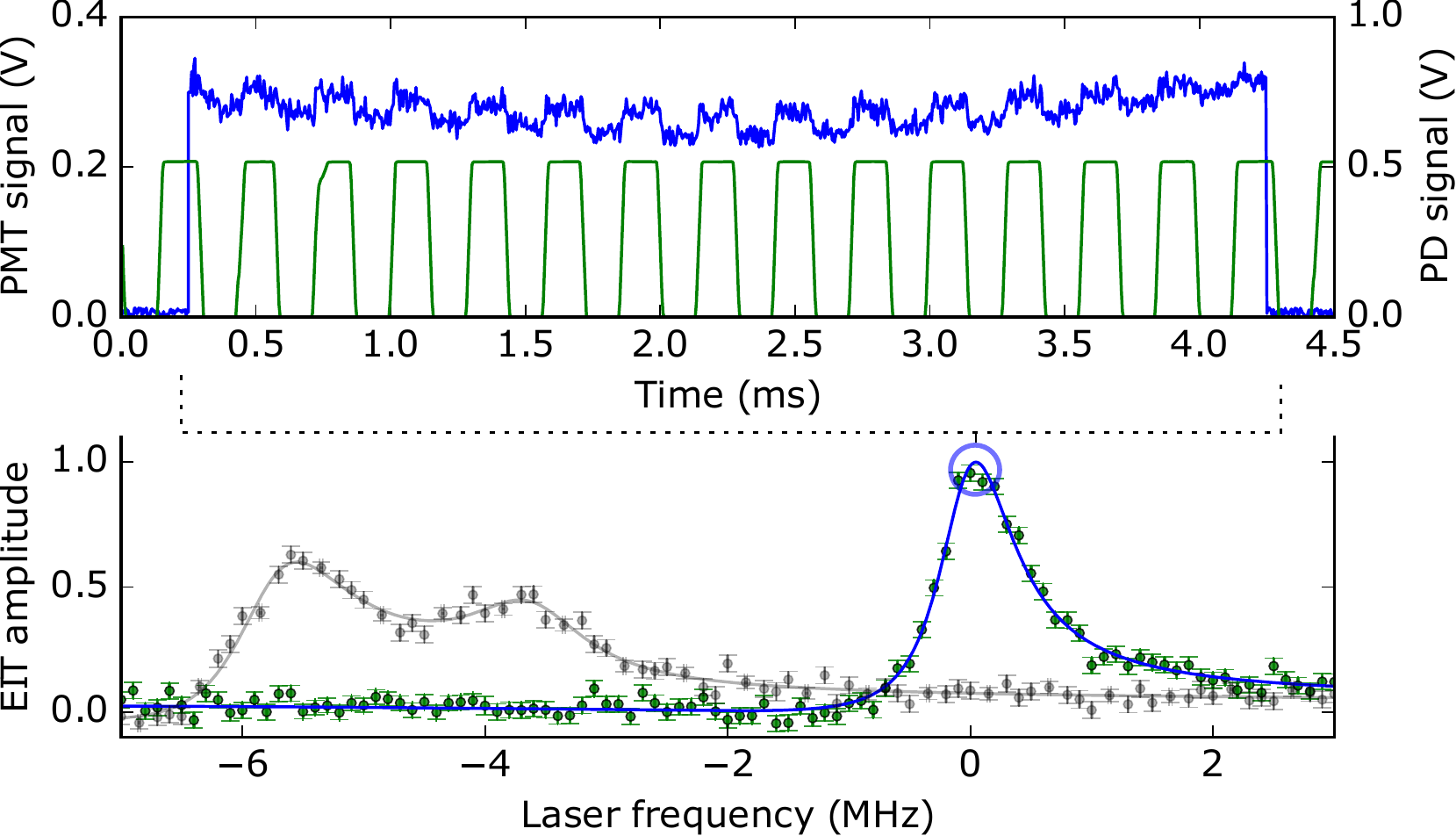}
    \caption{\label{fig:timetrace} \textit{top:} Modulation of the pump beam intensity (green) and its corresponding modulated EIT induced on the probe absorption signal (blue). The pump beam is resonant with the $n=53$ Rydberg state and both the external magnetic and electric fields set to zero. \textit{bottom:} An EIT spectrum is obtained by scanning the pump beam frequency and repeating such an absorption measurement. Example spectra are shown for zero applied magnetic field, with the applied electric field on (grey) and off (blue).  }
\end{figure}

For the implementation of the EIT spectroscopy counter-propagating beams, one resonant with the $5s^2$~$^1$S$_0-5s5p$~$^1$P$_1$ transition at 461~nm and the other with tunable frequency at 413 nm, excite atoms to a chosen 5$snd$~$^1$D$_2$ Rydberg state. The resulting EIT signal is measured using `lock in' detection of the 461~nm probe beam absorption via modulation of the 413 nm pump beam intensity by an optical chopper. A typical absorption measurement showing the modulated EIT signal induced by the pump beam is shown in the top of Fig.\ref{fig:timetrace}. The probe beam is derived from a commercial frequency doubled diode laser system. Its power and waist, as defined by the 1/$e^2$ radius of the intensity profile, are 800~fW and 120 $\upmu$m, respectively. At this power and atomic number, the probe beam absorption is between $20 - 40\%$. A home built extended-cavity diode laser (ECDL) provides 8 mW of pump light which is focused to an 80 $\upmu$m waist at the atoms' position. The bottom inset in Fig.\ref{fig:timetrace} shows  typical spectra taken at zero magnetic field with and without the applied external electric field.

%

The long term frequency stability of the pump and probe beam is maintained to within 10 kHz by locking to a transfer cavity referenced to the `clock' laser. In the case of the probe beam, the sub-harmonic at 922 nm is directly locked to the cavity. To stabilize the 413~nm pump laser frequency, a commercial Ti-sapphire laser, which is typically used to form the magic wavelength lattice at 813~nm, is first tuned to 826 nm and locked to the transfer cavity. This light is then frequency doubled by a LBO crystal \footnote{Purchased from Eksma Optics, Optolita UAB Mokslininku str. 11 LT-08412 Vilnius Lithuania} to produce 20 $\upmu$W as needed to generate a beat-note with the pump beam. The beat-note signal is mixed with a direct digital synthesiser (DDS) and the intermediate frequency is stabilized using a delay line offset lock scheme \cite{Schunemann1999} via fast feedback to the diode current and slow feedback to the ECDL piezo. The DDS provides the necessary tunability needed for scanning the pump laser frequency.  

In order to spectroscopically resolve the Zeeman sub-levels of the Rydberg state, an external magnetic field between 100-300 $\upmu$T is applied orthogonal to the propagation direction of the pump and probe beams. In this low field regime, the Zeeman splitting of the intermediate $5s5p$~$^1$P$_1$ state is negligible compared to its natural linewidth. The probe beam is linearly polarised orthogonal to the quantization axis to enable a balanced access to all the $m_J$ levels within the Rydberg manifold given the fixed polarisation of the pump light. Any background residual magnetic field is nulled using electron-shelving spectroscopy on the narrow-linewidth $5s^2$~$^1$S$_0-5s5p$~$^3$P$_1$ transition at 689 nm \cite{Akatsuka2008}. We resolve a Doppler-broadened linewidth of 40 kHz which constrains any residual field to below 2 $\upmu$T. 

To test the sensitivity of our method, an external plate electrode located directly opposite a radial DN40 viewport is used to apply a DC electric field in order to induce a Stark shift of the Rydberg states. Shielding from the metal vacuum chamber greatly attenuates the applied field at the atoms' position, meaning several kV potentials are needed to induce a substantial Stark shift. Such large potentials have the unfortunate effect of charging the dielectric viewport resulting in an exponential decay of the applied electric field strength at the atoms' position as inferred from a reduction in the Stark shift with time. To ameliorate this effect, we interleave measurements with opposite field polarity to avoid charging any external surfaces.

\section{ Rydberg Electrometry using EIT Spectroscopy }
\begin{figure}[h]
    \centering
    \includegraphics[width=0.48\textwidth]{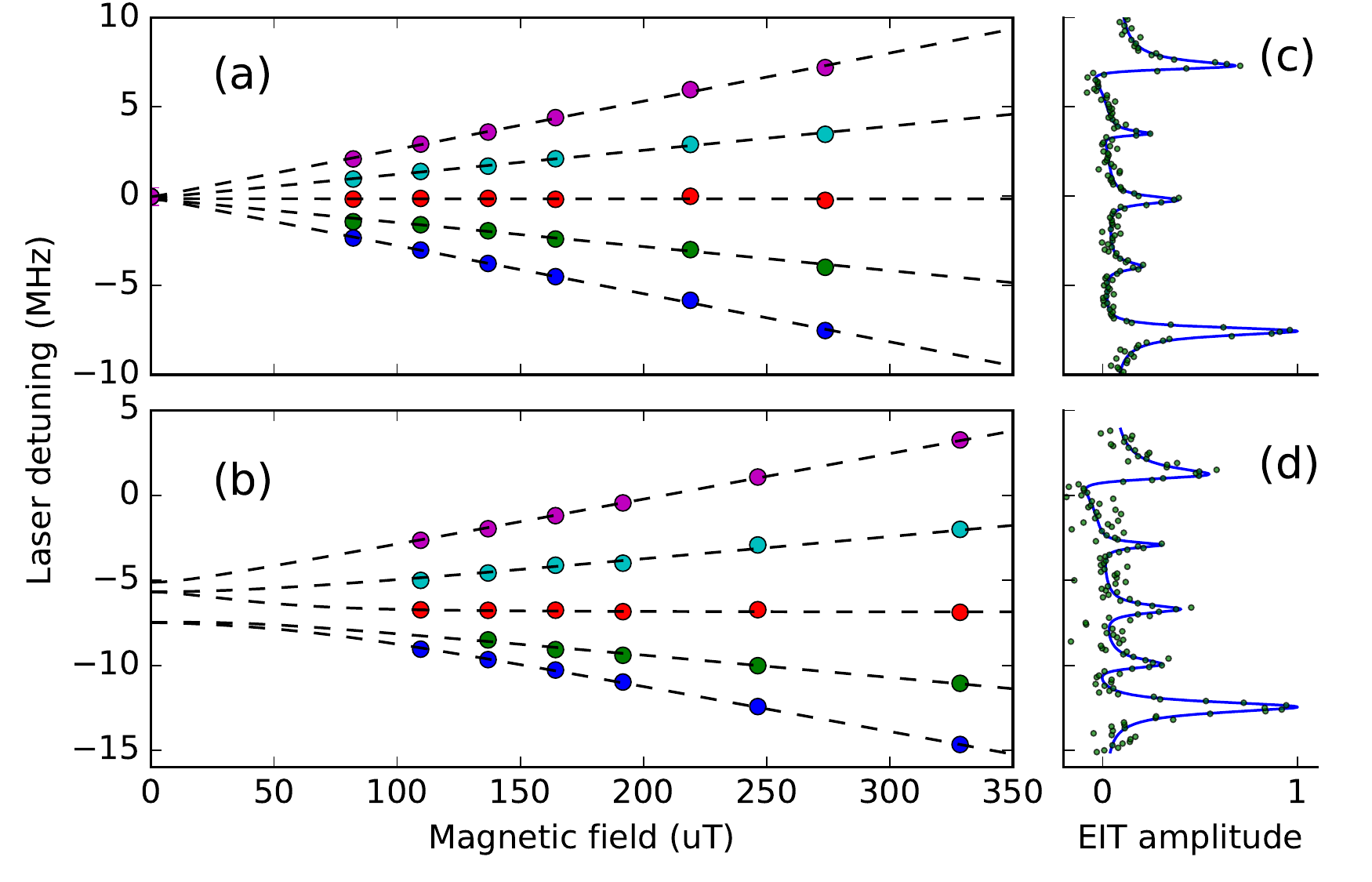}
    \caption{Inset $a$ shows the Zeeman splitting without an applied electric field for the Rydberg state n = 75 (the error bars are smaller than the symbols). The line centers are extracted from the EIT spectra, an example of which is shown in inset $c$ for the highest field case. Even without the an applied field, a fit to the splitting reveals a slight asymmetry resulting from the tensor nature of the Stark shift consistent with a residual electric field of 1.52~$\textrm{V} \textrm{m}^{-1}$. For the applied field case, shown in insets $(b)$ and $(d)$, the Stark shift is clearly visible and the fitting procedure returns an electric field of 5.75$~\textrm{V} \textrm{m}^{-1}$}
    \label{fig:Efit}
\end{figure}

As we do not have a method to directly measure the pump and probe frequencies, we have instead developed a method for measuring the applied electric fields that is based on the relative splitting of spectral lines, rather than the absolute detuning. In the absence of an electric field, the Zeeman sublevels split symmetrically with applied magnetic field \textbf{\textsl{B}}. However, in the presence of an electric field \textbf{\textsl{E}}, the Stark shift will result in an asymmetry in the spectrum, since the Stark shift depends on $|m_\mathrm{J}|$. Example spectra with and without an applied electric field are shown in Fig.\ref{fig:Efit}. In order to extract the line centers, five Fano profiles are fit to each spectrum corresponding to each of the five possible $m_J$ transitions. The asymmetric effect of an applied electric field on the observed Zeeman shift of the $m_\mathrm{J}$ is clearly visible in Fig.\ref{fig:Efit}.

To obtain the electric field from the spectroscopic data, the relative line positions are compared to a calculation of the combined Zeeman and Stark shift of each level. In the general case, the magnetic and electric field vectors are separated by an angle $\beta$, requiring transformation to a common basis. Choosing to work in the $\ket{J, m}$ basis defined by the magnetic field quantization axis, the matrix elements of the Zeeman Hamiltonian are give by

\begin{equation}
\bra{J, m_1}H_B\ket{J,m_2}=-m_1\upmu B \delta_{m_1m_2}
\end{equation}

\noindent where $\mu$, the magnitude of the magnetic dipole moment, is the Bohr magneton for a singlet state and $\delta$ is the Kronecker delta function. The Stark Hamiltonian, with eigenenergies $\Delta_E(m_J, E)$ that are computed as outlined in section \ref{theory},  is rotated by an angle $\beta$ by applying the appropriate Wigner D-matrix, $d^J_{m,m'}(\beta)$ for $J$ = 2. The matrix elements of this transformed Hamiltonian are given by

\begin{equation}
\bra{J, m_1}H_E\ket{J,m_2}= \sum_{m'} d^J_{m_1,m'}(\beta)d^J_{m_2,m'}(\beta)\Delta_E(m', E) 
\end{equation}

\noindent Finally, the theoretical splitting is computed by diagonalizing the Hamiltonian $H = H_{\textrm{E}} + H_{\textrm{B}}$. Using this approach, we fit the experimentally observed energy splitting by varying the electric field strength \textit{E} and its angle $\beta$ relative to the applied magnetic field in the model.  As the Stark shift is quadratic, our method only determines $\beta$ modulo $\pi$. The only other fitting parameter is an overall two photon detuning from the zero field resonance as we have no measure of the absolute frequency of the 413~nm laser. 

Fig.\ref{fig:Efit} shows the relative energy splitting for various magnetic fields with and without an applied electric field. An external electrode set to 2 kV, the maximum allowed by the high voltage supply, generated the applied electric field.  From a fit to the Zeeman splitting, the electric field at the position of the atoms is estimated to be $5.75\pm 0.11(\textrm{stat})\pm 0.16 (\textrm{sys}) \textrm{V} \textrm{m}^{-1}$. The fitting procedure also returned a value of $\beta = 0.47(1)\pi$ that is consistent with the axial magnetic field and radially applied electric field. An electric field of such magnitude would result in a fractional frequency shift of the clock transition equal to $2\times10^{-19}$. Given this is the largest field we can apply, it would have taken approximately a year of continuous operation to resolve this frequency shift given our fractional frequency instability, highlighting the utility of this method when applying large external fields is not possible. 

Next the external field was switched off and the procedure was repeated. A fit to the resulting splitting revealed a residual electric field of 
$1.52^{+0.62(\textrm{stat})}_{-.22}$$^{+0.05(\textrm{sys})}_{-.03} \textrm{V} \textrm{m}^{-1}$ most likely due to patch potential on the surrounding chamber. The uncertainty for this electric field value is comprised of both statistical error resulting from the fitting procedure and systematic error arising the uncertainty in the Stark map for $75^1$D$_2$. The quoted statistical error corresponds to a $68\%$ confidence interval as determined by the fitting procedure. A weak correlation observed between the uncertainty in $\beta$ and the electric field is taken into account in this estimate \cite{Press1988}, see appendix \ref{fitting} for further details. The systematic error on the electric field value due to the uncertainty for the Rydberg polarizability  was calculated by repeating the fitting procedure with revised Stark maps offset from the theoretically predicted value by $\pm\sigma$. Translating this electric field and corresponding uncertainty to the DC Stark shift of the $^1$S$_0$-$^3$P$_0$ clock transition results in a fractional frequency shift of $-1.6^{+0.4}_{-1.6}\times10^{-20}$. The fractional uncertainty of the differential polarizability of the clock states is negligible compared to that of the electric field and therefore has been ignored in the quoted uncertainty.

\section{Conclusion}
In conclusion, we believe that Rydberg electrometry constitutes a valuable technique for controlling systematic errors in optical lattice clocks. DC Stark shifts can in principle be separated from other systematic uncertainties using measurements on the clock transition alone, but this is time-consuming and requires the application of well-characterized external electric fields. In contrast, Rydberg states selectively enhance the spectroscopic sensitivity to stray electric fields by several orders of magnitude. The high spectroscopic resolution provided by EIT thus enables rapid quantitative measurements of the stray electric field. On the practical side, all that is required is a single additional laser to provide the pump beam, and existing lattice clock setups need not be modified to include electrodes. The constraint on the clock uncertainty that we obtained is compatible with the accuracy of the current generation of lattice clocks, and improved spectroscopy of the relevant Rydberg levels would see this reduced to negligible levels. Lastly, we note that the combination of Rydberg states and optical lattice clocks could also be applied to measurements of blackbody-induced systematic errors, and the creation of non-classical states.

\section{Acknowledgments}

The authors would like to thank Marco Schioppo for careful reading of our manuscript and Elizabeth Bridge for useful discussion regarding the feasibility of strontium Rydberg atoms for electrometry. This work was done under the auspices of UK NMO program. WB would like to acknowledge the support from the Marie Curie Initial Training Network FACT. PH and MPAJ acknowledge C Vaillant for useful discussions and the early version of the Stark map code. The experimental Stark map in Fig 1 was measured by R Hanley. They also acknowledge financial support from EPSRC grant
EP/J007021/, and EU grants FP7-ICT-2013-612862-HAIRS, H2020- FETPROACT- 2014-640378-RYSQ and H2020-MSCA-IF-2014-660028.

\appendix
\section{theoretical stark map uncertainty}

Here we consider further the estimation of the uncertainty in the electric field that arises from the calculation of the Stark map. As stated earlier, at the current level of precision the dominant contribution is the uncertainty in the experimentally-measured zero-field energies of the Rydberg states.
In this work, we carried out a more detailed analysis of the data set discussed in \cite{Vaillant2012}.
For each Rydberg series, the experimentally available Rydberg energies are fitted with the Rydberg-Ritz formula to obtain the zero-field  energy at all principal quantum numbers. Given that the reduced $\chi^2$ of those fits is around 2 with around 30 degrees of freedom, the statistical uncertainty on the fit parameters obtained from a simple fitting procedure can be underestimated \cite[p.107]{hughes2010}. To get a more reliable estimate of the uncertainty on the fit parameters, we fitted a large number of different sub-samples of the spectroscopic data, generated by randomly removing a few data points from the full data set. The mean value and uncertainty of the fit parameters are then taken to be respectively the mean and standard deviation $\sigma$ of the resulting distributions.

To obtain the uncertainty in the Stark map, the zero-field energies are varied within $\pm\sigma$ of their mean value. By understanding how each state shifts with applied field, we are able to find the combination of zero-field uncertainties that leads to the minimal and maximal values of the Stark shift, giving the curves that delimit the shaded area in Fig.1 in the article. 

\section{fitting procedure for Rydberg Zeeman splitting}
\label{fitting}
Determining the magnitude of the electric field from the Rydberg electromagnetic induced transparency (EIT) spectra requires a two step fitting procedure. First each EIT spectrum, recorded at a fixed magnetic field, is fit to determine the line centers of the five $m_J$ magnetic sublevels of the $5s75d$~$^1D_2$ Rydberg State. Using these values, the entire Zeeman splitting is then fit to extract both the electric field strength $E$ and the angle $\beta$ between the electric and magnetic field. 

To fit the EIT signal $T$ as a function of laser detuning $\Delta$, we use the following model:

\begin{equation}
T (\Delta) = \sum_{i=0}^{5} A_i(1-\frac{(q_i\Gamma_i+\Delta-\Delta_{0,i})^2}{\Gamma_i^2 + (\Delta-\Delta_{0,i})^2})
\end{equation}

\begin{figure}[h]
    \centering
    \includegraphics[width=0.5\textwidth]{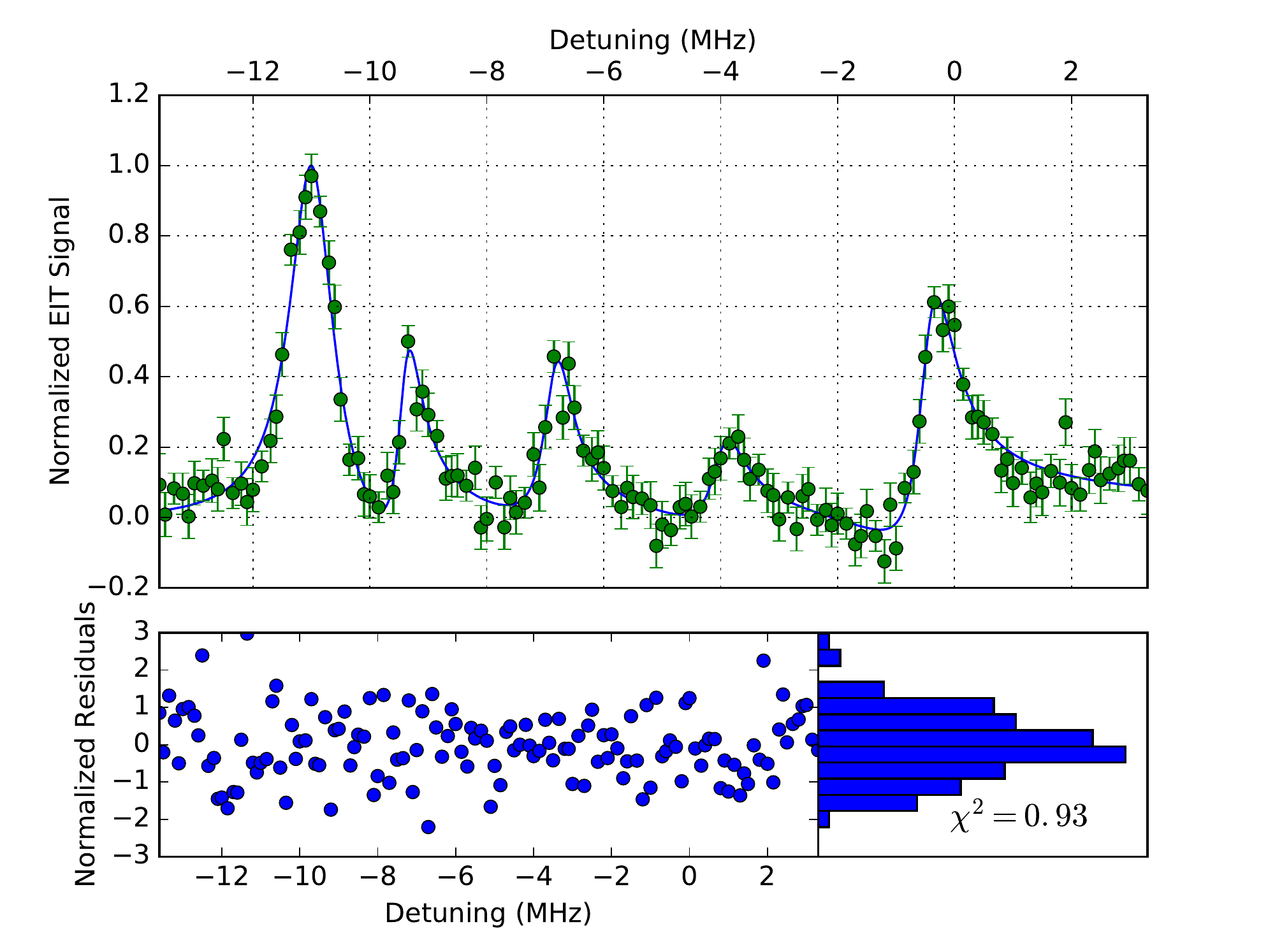}
    \caption{Example Rydberg spectrum taken at a magnetic field of .192 mT. The model shows good agreement with the observed data as indicated by the normally distributed residuals and the reduced $\chi^2$ value.}
    \label{fig:RydbergFit}
\end{figure}

\noindent The model is composed of five Fano profiles each of which has four fitting parameters: amplitude $A$, the linewidth $\Gamma$, the line centre $\Delta_{0}$ and the Fano parameter $q$. Fig.\ref{fig:RydbergFit} shows a fit along with the normally distributed residuals.

Once all spectra are fit, the observed Zeeman splitting is compared to the eigenvalues of the Hamiltonian containing both the Zeeman and Stark sub-terms. To compute this Hamiltonian, we adopt the notation that $\ket{J, m}$ and $\ket{J, m'}$ are the eigenstates of the Zeeman and Stark sub-terms, respectively. Transforming between these two basis amounts to a rotation by an angle $\beta$ represented by a operator $R=e^{-i\beta J_y}$. The matrix elements of the Stark Hamiltonian $H_E$, with eigenvalues $\Delta_E(m', E)$, in the $\ket{J, m_J}$ basis are given by:

\begin{equation}
\begin{split}
\bra{J, m_1}&H_E\ket{J, m_2}= \\
&\quad\bra{J, m_1}R\sum_{m'} \Delta_E(m', E)\ket{J, m'}\bra{J, m'}R^{\dagger}\ket{J,m_2}\\
&= \sum_{m'} \bra{J, m_1}R^{\dagger}\ket{J, m'}\bra{J, m'}R \ket{J,m_2}\Delta_E(m', E)\\
&\triangleq \sum_{m'} d^J_{m_1,m'}(\beta)d^J_{m_2,m'}(\beta)\Delta_E(m', E)
\end{split}
\end{equation}

\noindent where $d^J_{m_1,m'}(\beta)$ is the Wigner $d$-matrix for $J = 2$. In the final line, we used the identity $d^J_{m,m'}(\beta)=d^J_{m',m}(-\beta)$. By varying the values of the electric field strength and angle $\beta$, we find those that give the best agreement with the observed Zeeman splitting. Also included as a fitting parameter is an overall two photon detuning as we have no measure of the absolute frequency of the 413~nm laser.

\begin{figure}[h]
\includegraphics[width=.38\textwidth]{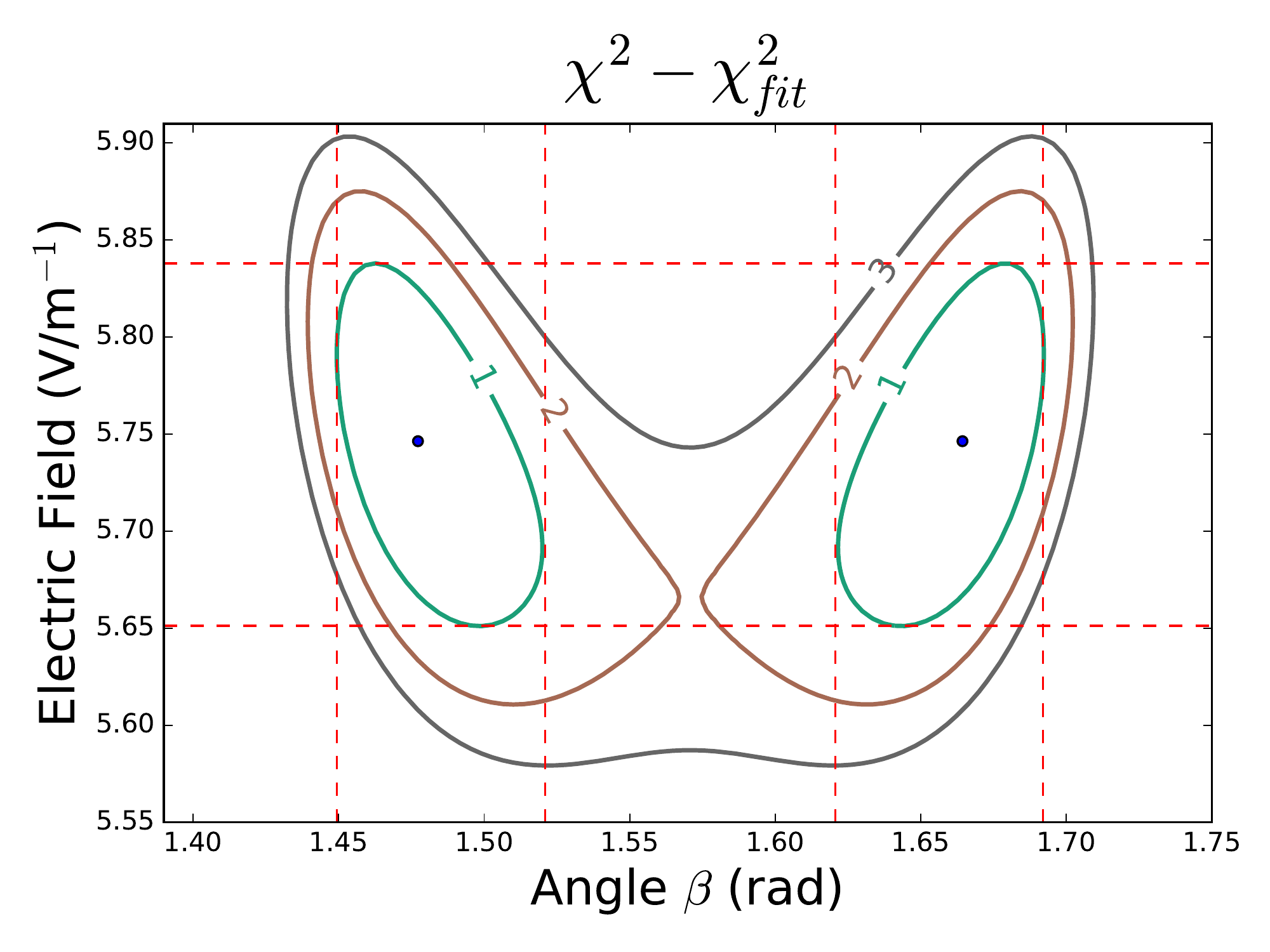}
\includegraphics[width=.38\textwidth]{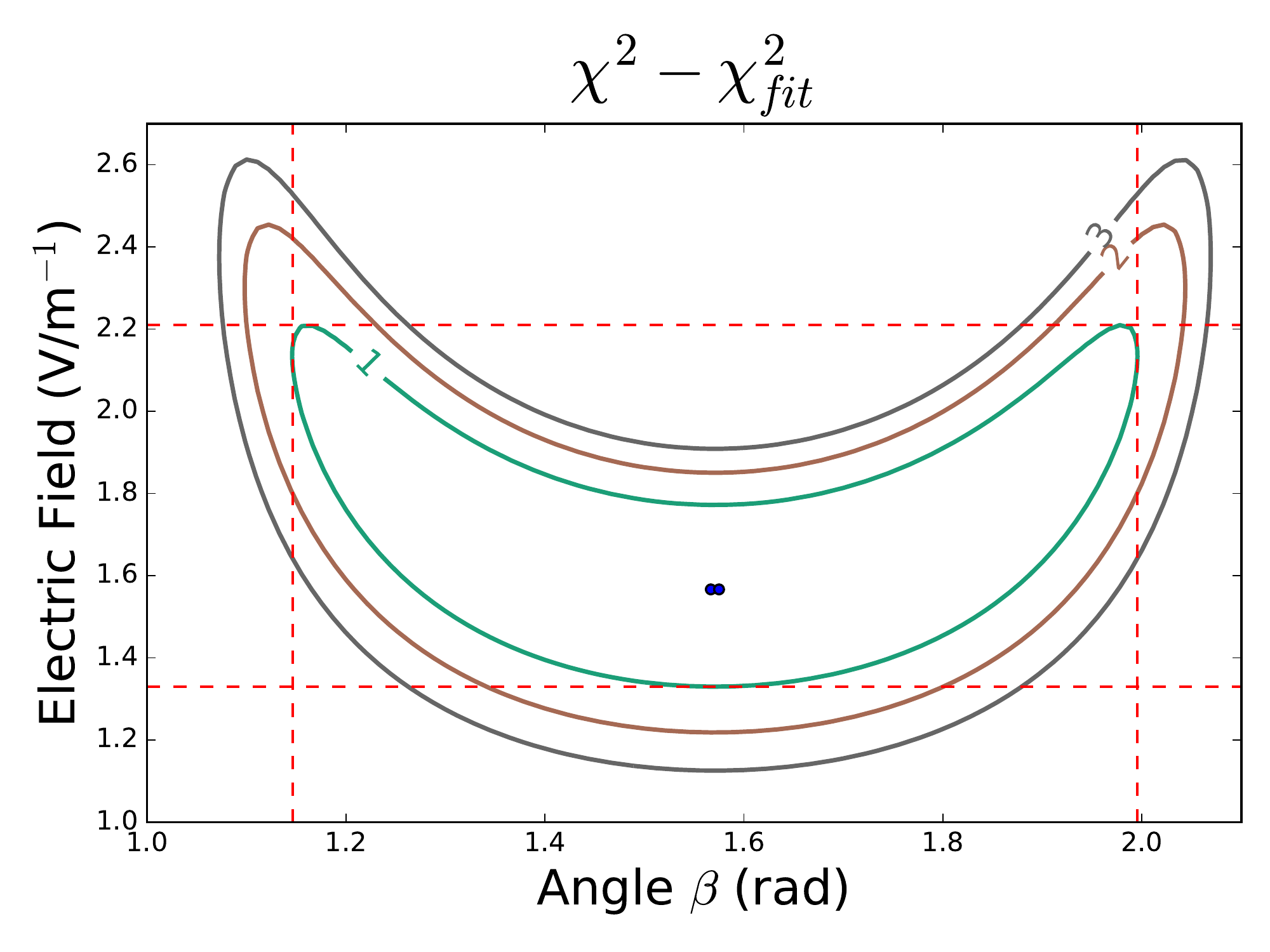}
\caption{Contour plots showing the change in $\chi^2$ resulting from varying the electric field strength and angle $\beta$ in the fitting model. Contours are shown for the datasets with(\textit{top}) and without(\textit{bottom}) an externally applied electric field. From the contours we can determine the uncertainty for the fitting parameters along with correlation between their values. The blue dots indicates the value for $E$ and $\beta$ that minimize $\chi^2$. The red dashed lines indicate the $\pm1\sigma$ bounds for each fitting parameter based on the extrema of the contour corresponding to increase in $\chi^2$ of 1.}
	\label{fig:contour}
	
\end{figure}

To extract the uncertainty for the electric field and angle $\beta$, we examine the sensitivity of $\chi^2$ to changes of their values. To do this we fix the their values near those that minimize $\chi^2$ and refit the data by only varying the two photon detuning. By repeating this procedure for different electric field and $\beta$ values, we produce a contour plot showing the change in $\chi^2$ as function of these two fitting parameters. The region bounded by the contour corresponding to an increase in $\chi^2$ of 1 sets the confidence region for the two fit parameters. The extrema of this boundary defines the quoted $\pm1\sigma$ uncertainties. Examples of such contour plots for the fits presented in this article are shown in  Fig.\ref{fig:contour}. 

This procedure also reveals correlations between the two fitting parameters that cannot be detected by the numerically computed covariance matrix. These correlations give rise to the asymmetric statistical errors quoted in the article. One can also see that $\chi^2$ is symmetric about $\beta = \frac{\pi}{2}$. This results from the quadratic nature of the Stark shift. Therefore, our method returns the value of $\beta$ modulo $\pi$. The numerical fitting package used for this work was the LMFIT: Non-Linear Least-Square Minimization and Curve-Fitting developed for Python \cite{newville_2014}.
\vfill\eject

\bibliography{myBib}

\end{document}